# A specifically designed machine learning algorithm for GNSS position time series prediction and its applications in outlier and anomaly detection and earthquake prediction

M. Kiani

**Abstract**. We present a simple yet efficient supervised machine learning algorithm that is designed for the GNSS position time series prediction. This algorithm has four steps. First, the mean value of the time series is subtracted from it. Second, the trends in the time series are removed. Third, wavelets are used to separate the high and low frequencies. And fourth, a number of frequencies are derived and used for finding the weights between the hidden and the output layers, using the product of the identity and sine and cosine functions. The role of the observation precision is taken into account in this algorithm. In order to test the algorithm and compare its performance with other machine learning methods, a large-scale study of three thousand position times series of GNSS stations across the globe is presented. Seventeen different machine learning algorithms are examined. The accuracy levels of these algorithms are checked against the rigorous statistical method of Theta. It is shown that the most accurate machine learning algorithm is the method we present. Furthermore, the method is at least 3 times faster than other methods. Two applications of the algorithm are presented. In the first application, it is shown that the outliers and anomalies in a time series can be detected and removed by the proposed algorithm. A study is presented for two thousand real time series, with ten thousand simulated outliers. Ten other methods of time series outlier detection are compared with the proposed algorithm. The study reveals that the proposed algorithm is approximately 3.22 percent more accurate in detecting outliers. In the second application, the suitability of the algorithm for earthquake prediction is investigated. A case study is presented for the Tohoku 2011 earthquake. It is shown that this earthquake could have been predicted approximately 2 hours before its happening, solely based on each of the 845 GEONET station time series. Comparison with four different studies show the improvement in prediction of the time of the earthquake.

―――――――

**Introduction**

Machine learning algorithms deal with finding the hidden information in data in an automated manner (Hastie et al. 2008). The data that is used in these algorithms are of different type. However, time series-values of a function at different times-is one of the most commonly used inputs to the machine learning algorithms (Hastie et al. 2008), (Ahmed et al. 2010). These algorithms are usually classified into three main groups: supervised, unsupervised, and reinforcement learning (Orr 1996). In these categories, the supervised case is of particular interest in many fields, such as economics and stock market (Makridakis et al. 2018), robotics (Tatinati 2015), electrical engineering (Dong et al. 2017), sustainable development (Garg et al. 2019), and alike. Problems such as classification, prediction, and regime switching are studied in various fields, using machine learning techniques (Orr 1996).

Different machine learning algorithms are designed and used for the purpose of time series analysis, including multilayer perceptron, Bayesian regulation, Gaussian processes and alike (Alpaydin 2014). However, these algorithms are purely mathematical and do not consider the fundamental, unique characteristics of the time series to which they are applied. For instance, the observation type and its accuracy are not taken into account. This would not have much effect on the processes involving data from fields such as economics, because the nature of the observations, which is usually not under the physical conditions of the environment, does not require weighing the observations. However, this is not always true, particularly for the observations gathered by instruments that are under various environmental conditions. One such case is the GNSS position time series. This type of time series includes the position of a station at different times (Blewitt et al. 2018).

GNSS position time series display certain fundamental characteristics that seem necessary to be considered for any possible algorithm to be devised. These include: the low value of the ratio of changes in coordinates to the coordinate themselves (excluding special circumstances such as earthquake), the normal presence of an ascending or descending trend, and presence of high and low frequencies in the data. These are usually neglected in the conventional machine learning algorithms.

One of the problems pursued by means of machine learning algorithms and that is of much interest in different fields is the prediction problem. Unlike the usual interpolation problems such as splines (Kiani and Chegini 2019), (Kiani 2020a), (Kiani 2020b), moving least squares (Kiani 2020c), (Kiani 2020d), remote sensing smoothing and classification tools (Kiani 2020e), (Kiani 2020f), (Kiani 2020g), the prediction problem is indeed an extrapolation, which is normally called propagation, because of the auto-regressive nature of the problem (Alpaydin 2014). The situation here is much like the orbit propagation by numerical integration (Kiani 2020h). Prediction in the field of geodesy is of much interest and use, possibly in hazard prediction assessment, such as the case with earthquakes and land subsidence and upheave prediction (Kiani 2020i), (Kiani 2020j), (Kiani 2020k). However, as it was stated earlier, the present conventional machine learning algorithms do not consider the characteristics of the GNSS time series and thus, are unsuitable for the prediction problem. We address this problem: devising a new algorithm for the GNSS position time series prediction is the goal of the present paper. Based on this, the major contributions of the present paper can be summarized as the following.

- Devising a new algorithm for the time series prediction that is directly based on the fundamental characteristics of the GNSS position time series

- Demonstrating the superiority of the proposed algorithm by checking its performance against the well-established machine learning algorithms, in a large-scale study of thousands of time series

- Presenting the applications of the proposed algorithm in the field of geodetic science and geodesy, particularly for outlier and anomaly detection, and earthquake prediction

The rest of the paper is organized as follows. In section 2, the mathematical basis for the algorithm is developed. In section 3, a large-scale study for the relative performance evaluation of the algorithm is presented. Section 4 is devoted to the applications of the proposed algorithm. Finally, section 5 is used for expressing the conclusions.

**Explanation of the algorithm**

The algorithm is based on simple mathematical manipulations, chosen according to the characteristics of the GNSS position time series, mentioned in the previous section. Note, however that in the first characteristic mentioned the changes in GNSS time series are not completely periodic but sometimes increase or decrease in amplitude, which we thence refer to as semi-periodic changes.

The algorithm is a supervised learning and thus is decomposed into two phases: training and prediction. The inputs of the machine learning algorithm-in the training phase-are the time, $t_k$ (different values for the index $k$ represent different times), and values of the time series at these times, $y_k$. The prediction problem is defined as follows.

**Definition 1.** The auto-regressive prediction problem is the determination of $y_{k+1}$ based on the $n$ training data $y_k, y_{k-1}, \ldots, y_{k-n+1}$, using the function g in the hidden layer that connects the inputs and outputs, as the following

$$y_{k+1} = g(y_k, y_{k-1}, \ldots, y_{k-n+1}) \tag{1}$$

The function g is what distinguishes the many different machine learning algorithms from each other. Based on the second and third characteristics of the GNSS position time series mentioned, the explicit form of the function g is given in the following.

**Definition 2.** The explicit form of the trend and semi-periodic terms of the function g, in Definition 1, denoted thence by $g_T$ and $g_P$ and applied to the input data at separate stages, are as the following

$$g_T(t) = a + bt$$

$$g_P(t) = \sum_{k=1}^{m} c_k t\cos(2\pi f k t) + s_k t\sin(2\pi f k t) \tag{2}$$

where $a, b, c_k, s_k$ are the weights that must be determined based on inputs, and $m$ is the degree used for the approximation. The $f_k, k = 1, \ldots, m$, however, are the frequencies in the input data. We determine these values as the following.

**Definition 3.** The frequencies in Definition 2 are calculated based on the fundamental frequency $f_0$ and the value of m as follows

$$df = \frac{f_0}{p+2}$$

$$f_1 = f_0$$

$$f_k = f_0 - (k-1)df, \ k = 2, \ldots, m \tag{3}$$

where p is the least power of 2 which is greater than n. For instance, n = 214 inputs corresponds to p = 256.

**Remark 1.** The fundamental frequency in this paper plays an important role. This value is based on the sampling frequency of the time series, usually $\frac{1}{86400}$ Hz. However, high-resolution time series, such as the ones presented in section 4, have the sampling rate of 1 second, meaning $f_0 = 1Hz$.

With these preliminaries, the steps in the algorithm are explained in the following.

**Step 1.** In this step, the average value of the inputs $y_k, y_{k-1}, \ldots, y_{k-n+1}$ is computed and subsequently subtracted from the time series.

$$M = \frac{1}{n}\sum_{j=0}^{n-1} y_{k-j}$$

$$y_{k-l} = y_{k-l} - M, \ l = 0, \ldots, n-1 \tag{4}$$

This step is done to achieve smaller values, which then will be used for modeling by the mathematical functions. This biggest part is subtracted in this step but will be added in the prediction step.

**Step 2.** The trends are removed in this step. Instead of the least squares line-a line like the one in (2) to derive the coefficients of which all the training data are used-the line connecting the first and last point of the training data is used. This is done to avoid the effect of possible errors and anomalies in the time series. This means the trends in the time series are usually uniform and do not change in the time series much, except for anomalous observations. Therefore, the trends in the data in the previous step are removed as the following

$$a = -\frac{y_{k-n+1} - y_k}{t_{k-n+1} - t_k} \times t_k + y_k$$

$$b = \frac{y_{k-n+1} - y_k}{t_{k-n+1} - t_k}$$

$$y_{k-l} = y_{k-l} - a - bt_k, \quad l = 0, \ldots, n-1 \tag{5}$$

After this step, stationarity is achieved in the mean. Note that other methods exist for finding the trends in the data such as (Anghinoni et al. 2019). However, they are both more time consuming and not based on the characteristics of the GNSS time series.

**Step 3.** In order to find the prediction values more accurately, the high and low frequency components of the time series are separated and treated separately. To this end, the wavelets are used. If $y^l$ and $y^h$ denote, respectively, the low and high frequency components of the time series derived from applying a wavelet to the time series, one can write, based on (2)

$$y = y^l + y^h$$

$$y^l = \sum_{k=1}^{m} c_k^l t \cos(2\pi f_k t) + s_k^l t \sin(2\pi f_k t)$$

$$y^h = \sum_{k=1}^{m} c_k^h t \cos(2\pi f_k t) + s_k^h t \sin(2\pi f_k t) \tag{6}$$

in which the two sets of coefficients $c_k^l, s_k^l, \ k = 1, \ldots, m$ and $c_k^h, s_k^h, \ k = 1, \ldots, m$ are, respectively, the coefficients of the low and high frequency components of the time series. Each of these sets of values is determined in a least squares process by having the n input and 2m unknowns.

**Remark 2.** An important difference between the algorithm presented in this paper and other conventional machine learning algorithms is that the in the adjustment phase to compute $c_k^h, s_k^h$, $k = 1, \ldots, m$ and $c_k^h, s_k^h, \ k = 1, \ldots, m$ the observation's precision is also included in the processes. This would lead in a more realistic machine learning approach, in which the observations are treated with different weights. The simplest form of the weight of each observation, $w_k$, is as follows, with $\sigma_k$ being the accuracy of the $k$ th observation

$$w_k = \frac{1}{\sigma_k^2} \tag{7}$$

**Step 4.** In the last step, the prediction is made. The next value of the time series is predicted and subsequently used as an input for future predictions in the dynamic system of auto-regression. If $t_n$ and $y_n$ denote, respectively, the time at which the prediction is made and the value of prediction with regard to the system in (1), one can simply write

$$y_n^l = \sum_{k=1}^{m} c_k^l\, t_n \cos(2\pi f_k t_n) + s_k^l t_n \sin(2\pi f_k t_n)$$

$$y_n^h = \sum_{k=1}^{m} c_k^h\, t_n \cos(2\pi f_k t_n) + s_k^h t_n \sin(2\pi f_k t_n)$$

$$g_T(t_n) = a + b t_n$$

$$y_n = y_n^l + y_n^h + g_T(t_n) + M \tag{8}$$

**Remark 3.** It is important to notice that the training phase is different from the prediction phase. In fact, the proposed algorithm is first used for the training data themselves to achieve the best performance for the data. The performance is usually evaluated by the Mean Square Error (MSE). As long as the MSE is greater than a given threshold T, the number of frequencies changes. If $MSE < T$ the training phase is done, and m is fixed. Then the algorithm is used for the prediction of the next outcomes.

Figure 1 summarizes the proposed algorithm.

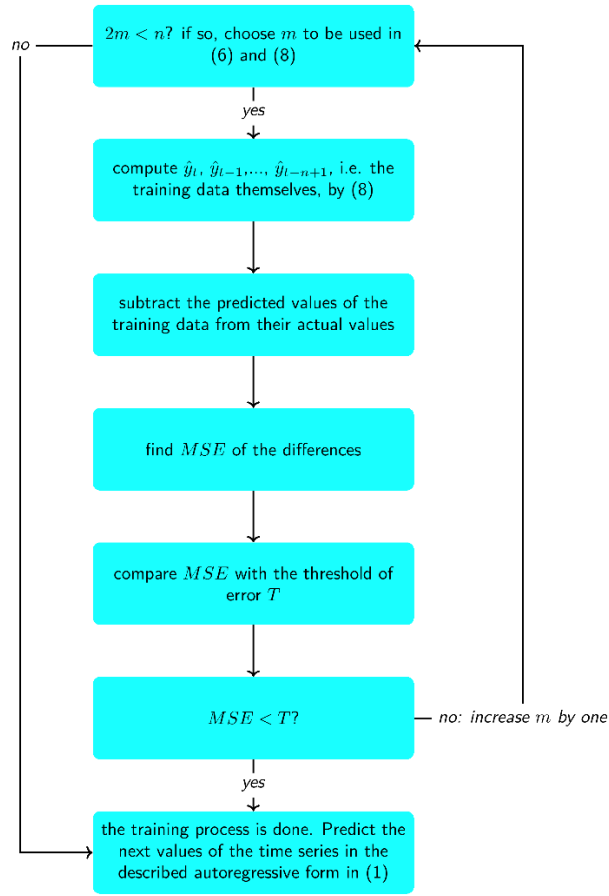

**Fig 1** The proposed algorithm

Performance evaluation

In order to assess the accuracy of the predicted values, different measures can be applied. To analyze the accuracy, one must have both the predicted and observed values, denoted by $y_i, i = 1, \ldots, q$ and $\hat{y}_i = 1, \ldots, q$, where q is the total number of predictions made. We use four different measures, namely, the symmetric Mean Absolute Percentage Error (sMAPE) (Ahmed et al. 2010), Mean Absolute Scaled Error (MASE) (Makridakis et al. 2018), Standard Deviation (StD), and Mean of Absolute Errors (MAE). These so-called error criteria are defined as the following

$$sMAPE = \frac{200}{q} \sum_{k=1}^{q} \frac{|y_k - \hat{y}_k|}{|y_k| + |\hat{y}_k|}$$

$$MASE = \frac{n-1}{q} \frac{\sum_{k=1}^{q}|y_k - \hat{y}_k|}{\sum_{j=2}^{Q}|y_j - y_{j-1}|}$$

$$StD = \sqrt{\frac{1}{q-1} \sum_{k=1}^{q} (E_k - \bar{E})}$$

$$MAE = \frac{1}{q} \sum_{k=1}^{q} |y_k - \hat{y}_k| \tag{9}$$

where $Q = n + q$ is the total number of data, $E_k = y_k - \hat{y}_k$, and $\bar{E} = \frac{1}{q}\sum_{k=1}^{q} E_k$.

sMAPE is expressed in terms of percentages and it is obvious the smaller its value the more accurate the predictions are. The criterion MASE is, as the name suggests, independent of the scale of the inputs.

**A large-scale study for the performance of the proposed method**

In this section, a large-scale study of 3000 GNSS position time series from stations all over the world for the evaluation of the performance of the proposed algorithm is presented. These time series are taken from Nevada Geodetic Laboratory (Blewitt et al. 2018), and are in IGS14 coordinate system. The mentioned four steps in the previous section are used for these time series. The average performance criteria are computed. In order to analyze the relative performance of the proposed method, seventeen different machine learning prediction algorithms are used, namely, multilayer perceptron (Watson 1964), Bayesian regulation (Kononenko 1989), BFGS quasi-Newton backpropagation (Robitaille et al. 1996), conjugate gradient backpropagation family (Powell-Beale, Fletcher-Reeves, Polak-Ribiere, and scaled), (Nawi et al. 2010), gradient descent backpropagation family (simple, adaptive learning rate, momentum, and momentum and adaptive learning rate), (Yu and Chen 1997), resilient backpropagation, (Mastorocostas 2004), radial basis functions network and its exact mode (Orr 1996), one-step secant backpropagation (Hagan and Menhaj, 1994), generalized regression (Specht, 1991), and batch training (Li et al. 2014). The Theta statistical method (Assimakopoulos and Nikolopoulos, 2000) is also used to compare the machine learning algorithms with the traditional statistical methods. To find the information about the algorithms the reader is referred to the mentioned references, and also to (Alpaydin, 2010). After evaluation of these algorithms, the following results have been obtained for the *X* component. Note that similar results exist for *Y* and *Z* components and thus we avoid mentioning them here.

**Table 1** Prediction accuracy of the 18 machine learning algorithms and one statistical method for 3000 different time series of GNSS stations, analysis for X component

| algorithm/method | sMAPE(%) | MASE | StD(m) | MAE(m) | Speed rank |
|---|---|---|---|---|---|
| proposed algorithm | $4\times10^{-6}$ | 0.934 | 0.022 | 0.031 | 1 |
| multilayer perceptron | $7\times10^{-6}$ | 47.878 | 0.043 | 0.069 | 16 |
| Bayesian regulation | $4\times10^{-6}$ | 26.225 | 0.030 | 0.050 | 13 |
| BFGS quasi-Newton backpropagation | $7\times10^{-6}$ | 38.604 | 0.043 | 0.071 | 15 |
| Powell-Beale conjugate gradient backpropagation | $7\times10^{-6}$ | 51.429 | 0.044 | 0.072 | 10 |
| Fletcher-Reeves conjugate gradient backpropagation | $7\times10^{-6}$ | 38.239 | 0.043 | 0.072 | 9 |
| Polak-Ribiere conjugate gradient backpropagation | $7\times10^{-6}$ | 50.458 | 0.042 | 0.072 | 12 |
| scaled conjugate gradient backpropagation | $1.4\times10^{-5}$ | 208.915 | 0.031 | 0.050 | 7 |
| simple gradient descent backpropagation | $4\times10^{-6}$ | 24.813 | 0.030 | 0.049 | 5 |
| adaptive learning rate gradient descent backpropagation | $1.4\times10^{-5}$ | 274.390 | 0.031 | 0.049 | 17 |
| momentum gradient descent backpropagation | $1.4\times10^{-5}$ | 36.315 | 0.031 | 0.049 | 4 |
| momentum and adaptive learning rate | $1.4\times10^{-5}$ | 34.756 | 0.031 | 0.050 | 18 |
| resilient backpropagation | $1.4\times10^{-5}$ | 34.779 | 0.032 | 0.051 | 6 |
| radial basis functions | $1.4\times10^{-5}$ | 33.248 | 0.031 | 0.049 | 19 |
| exact radial basis functions | $4\times10^{-6}$ | 24.769 | 0.030 | 0.050 | 3 |
| one-step secant backpropagation | $4\times10^{-6}$ | 45.647 | 0.030 | 0.049 | 14 |
| generalized regression | $4\times10^{-6}$ | 26.263 | 0.031 | 0.051 | 2 |
| batch training | $4\times10^{-6}$ | 23.429 | 0.030 | 0.050 | 8 |
| Theta statistical method | $7\times10^{-6}$ | 953.209 | 0.043 | 0.070 | 11 |

From the Table 1, as well as the similar results for the *X* and *Y* components, one can understand that the proposed method outperforms other conventional machine learning algorithms. In terms of sMAPE, it has the lowest values. Although some of the methods have comparable sMAPE values to that of the proposed algorithm, the more reliable MASE criterion (Makridakis et al. 2018) determines that the proposed algorithm is better than other methods, because only the MASE of this method is less than 1. Also, in stark contrast to the other methods' suitability in the time series prediction in other fields (Ahmed et al. 2010), they are not suitable for the special case of GNSS position time series. It can be understood, however, that the overall performance of the machine learning algorithms is better than the traditional, statistical Theta method. Comparison between the StD and MAE values in this table demonstrates that the best performance is achieved by the proposed algorithm. The StD values of the proposed algorithm are around 2 centimeters, while those of other methods are approximately two times larger. Also, considering the MAE values, one can conclude that the proposed algorithm has a better performance, by a scale of around two times. Hence, it can be said the proposed method has a better overall performance, compared to other conventional machine learning and statistical methods.

**Potential applications of the proposed algorithm**

In this section, two applications of the proposed algorithm in the field of geodesy and geodetic science are presented: outlier and anomaly detection in GNSS time series and earthquake prediction.

Outlier and anomaly detection in the time series

Outlier detection, or in another interpretation anomaly detection, is a very important topic in the field of geodetic science and many papers, including (Hekimoglu and Erenoglu 2007) and (Lehmann 2013) have analyzed this problem. In this section, however, we analyze this problem in time series. In this special case, the problem can be defined as finding outliers or anomalies in a time series by means of prediction and its comparison with the observed value. However, different problems may arise if the training data themselves contain outliers or anomalies. Indeed, this problem is complex and requires a specific strategy by means of which the problem of outliers in training can be overcome. For this purpose, we propose the following algorithm.

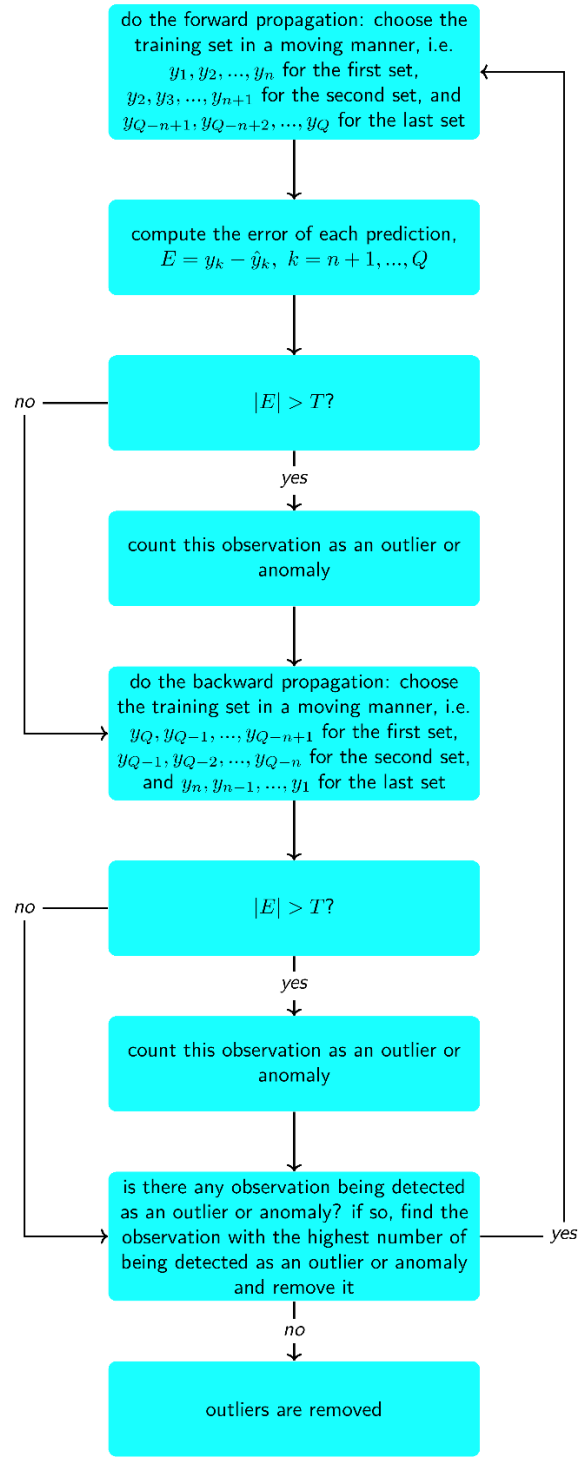

**Fig 2** proposed algorithm for anomaly and outlier detection

**Remark 4.** The smallest detectable error in the data depends on the time series and the range of its prediction errors. Factors such as the variability of the time series, amplitude of changes in the time series and availability of gaps in data all affect the prediction error. If the time series is not much variable and except for the outliers, it changes little, the amplitude of changes is also small- a few centimeters, and there are no gaps, we can expect that we could detect outliers with millimeter accuracy. Based on the discussion in the previous section, however, one can expect to detect anomalies or outliers as small as 3 centimeters, as a general estimation.

*A comparative analysis of the GNSS outlier detection success rate by different algorithms: results of simulation for 2000 time series.*

In order to analyze the success rate of the proposed algorithm in detecting the outliers, and compare it with those of other methods that are used for the purpose of outlier and anomaly detection in time series, the following algorithms are used. The first five are in used in specifically in geodetic science, while the next five are more general

- A1 Method in (Yakovlev, 2016)
- A2 Method in (Goudarzi et al. 2013)
- A3 Method in (Ogutcu, 2018)
- A4 Method in (Khodabandeh et al. 2012)
- A5 Method in (Wang et al. 2016)
- B1 Method in (Weekley et al. 2009)
- B2 Method in (Choy, 2001)
- B3 Method in (Hau and Tong, 1989)
- B4 Method in (Maya, 2019)
- B5 Method in (Yu et al. 2014)

In order to test the methods, a simulation is done. Two thousand time series are taken from (Blewitt et al. 2018), and 10000 errors with magnitude 2cm to 5m were included in the time series, at random time values in the time series. Each time series has at least 2 errors. Then the methods are applied to the resulting time series. In the following figure the simulation process is shown.

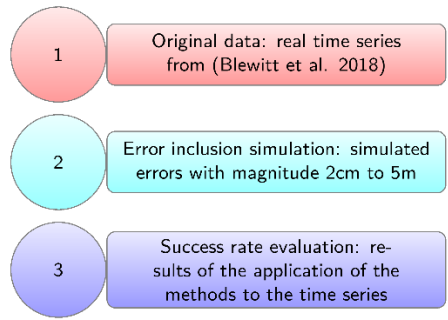

**Fig 3** The steps in success rate evaluation

Based on the diagram above, the following results are obtained in Table 2.

**Table 2** Evaluation of the success rate of the detection of outliers by the A1−A5 and B1-B5 methods and the proposed algorithm

| algorithm/method | number of detected outliers | success rate(percent) | rank |
|---|---|---|---|
| proposed algorithm | 9795 | 97.95 | 1 |
| A2 | 9473 | 94.73 | 2 |
| A4 | 9417 | 94.17 | 3 |
| B5 | 9297 | 92.97 | 4 |
| A5 | 9241 | 92.41 | 5 |
| B4 | 9155 | 91.55 | 6 |
| B2 | 8910 | 89.10 | 7 |
| A3 | 8888 | 88.88 | 8 |
| A1 | 8860 | 88.60 | 9 |
| B3 | 8812 | 88.12 | 10 |
| B1 | 8508 | 85.08 | 11 |

Based on Table 2, one can simply understand that the proposed algorithm is more capable of detecting errors and outliers. Its success rate is 3.22 percent higher than the next best method, which corresponds to revealing 322 more outliers. The reason is that in the method, data are analyzed from different times (i.e. forward and backward propagations), which result in detecting errors from different time.

Earthquake time prediction using position time series: a case study for the Tohoku 2011 earthquake

GNSS position time series can be of great use in areas with past records of earthquake and high seismological activity. Preseismic movements can be detected and used for the prediction of the earthquake. One such region is Japan, a country long known for its high risk of earthquakes. An example of this would be the Tohoku region in Japan, which was struck by a 9 Mw earthquake in 2011. Some studies have been devoted to this earthquake, including (Kawase 2014). However, in this paper we are primarily interested in showing the potential predictability of this earthquake. As it was shown in previous sections, the time series can be predicted by an accuracy of at least 3 centimeters. So, one would expect that such a large-magnitude earthquake as Tohoku could have been simply predicted. For this reason, the GEONET data in this region are taken from (Habboub 2019). These data contain 845 GPS position time series, before, during, and after the earthquake, in the local IGS00 datum denoted by $(E, N, U)$. The sampling rate of these data is 1 second. Data of the station number one is shown in Figure 4.

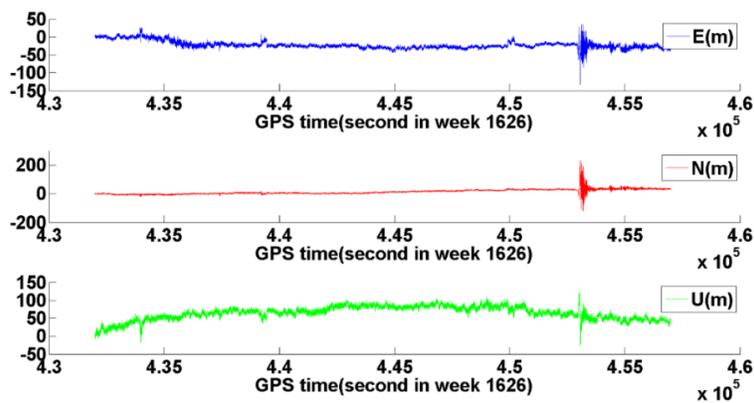

**Fig 4** Time series of the station number 1 in the GEONET

As it is seen from Figure 4, the effects of the earthquake start from around the second 452984. An important point about this time series is that the $E$ and $U$ components show the changes in coordinates better (it is true for all 845 time series), and we use these components.

It is also important to analyze the time series in a preliminary stage, to remove the effects of potential factors other than the ground movement that affect the preseismic values of the time series. Factors such as atmosphere and noise affect the time series data. So, using methodologies for removing atmospheric effects (Tregoning and Watson 2009) and GNSS time series noise and error contributions (Kaczmarek and Kontny 2018), (Ji et al 2020), (Williams et al 2004), time series is turned to a more realistic indicator of the ground movements. The following figure shows the steps to implement the proposed algorithm for earthquake prediction.

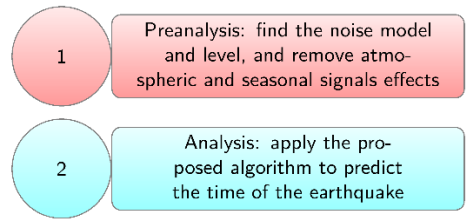

**Fig 5** The steps in earthquake time prediction

**Remark 5.** The number of training data is chosen in a way that the corresponding training data contain the first detected unusual movements. This means the difference between each point and its previous point is computed and if it is more than a given threshold, it would be considered an unusual movement, and this is the point of departure for analyses. Based on this, we set n = 17487 and m = 2000. With these values we get Figure 6.

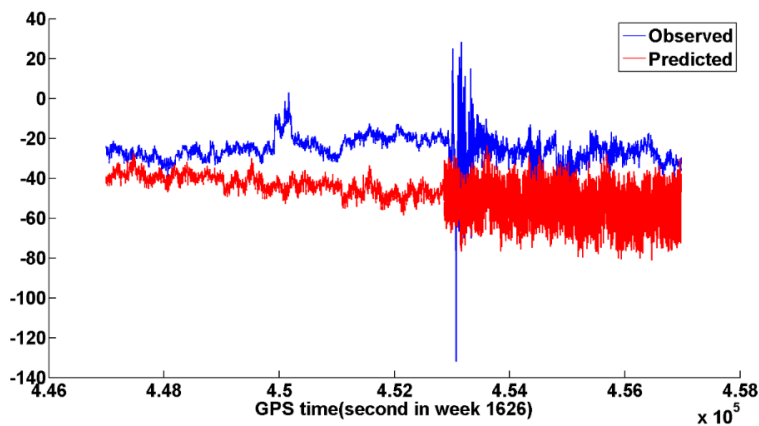

**Fig 6** Predicted earthquake in the station number 1 in the GEONET, *E* component

Note that although the prediction is not accurate for the movements after the earthquake, it predicts the time of the earthquake at around the second 452864 in week 1626, approximately 120 seconds before the actual time of the earthquake. The first amplitude of ground motion of the earthquake is predicted as 19.842 meters, i.e. as earthquake strikes, the $E$ coordinate changes 19.842 meters. The observed counterpart of this value is, however, 22.4 meters.

If the algorithm is section 2 is applied to these 845 time series, the average values for time predicted for the earthquake and the measure of ground motion are given in Table 3.

**Table 3** The average time and measure of the ground motion predicted for the Tohoku earthquake, across 845 stations

| time(second in GPS week 1626) | first ground motion (m) |
|---|---|
| 452807 | 18.539 |

*A quick comparison of the accuracy of the proposed algorithm with previous studies*

There are some studies related to the prediction of the Tohoku earthquake. Particularly (Li et al. 2013), (Liu and Zhou, 2012), (Psimoulis, 2013), and (Psimoulis et al. 2014) are of interest. In the first two studies, only a forewarning and precautionary analysis is done: there is no specific estimation of the time of the earthquake. In (Li et al. 2013) only unusual movements are detected and based on these, the risk of the earthquake happening is assessed. In (Liu and Zhou, 2012), only an estimation of the recurrence rate of earthquakes in northeastern Japan is presented. The combination of this data with the earthquake that happened two days before the main earthquake would have resulted in a high-risk estimate and earthquake premonition. As the paper asserts, however, this estimation based on the recurrence rates were not correct and the smaller earthquake was regarded as the main shock. So, seismological estimations were incorrect. However, the last two studies, new algorithms are described to predict the earthquake based on the preseismic data. Algorithms in these studies report the time of Tohoku earthquake approximately 12 seconds after its happening. This is a delayed estimation and comparing it with the estimated time in Table 3 is less accurate, since the latter is a forewarning and is estimated with a smaller number of data. Furthermore, we compensated for the effects of the atmosphere and noise of the signal in the pre-analysis phase. Besides, the amplitude of the first ground movement in the mentioned papers are not as accurate as the one in Table 3, since it is approximately the tenth of the size of the actual movement, which corresponds to an error of 90 percent. But according to Table 3, the error is almost 19 percent. Thus, the algorithm in this paper outperforms other methods for the estimation of time (and ground motion) of the Tohoku earthquake.

**Remark 6.** Using 60 percent of the data as the training data means the earthquake could be approximately predicted 2 hours 45 seconds before its happening. However, this is just based on the GPS time series; if other sources of information are combined with these data, it would result in a much better prediction. Furthermore, it can be simply tested that with fewer number of data, the predicted time of the earthquake is approximately the same, but the measure of the first ground movement is less accurate. For instance, using 10 percent of the data as the training data results in an estimation of the time of the earthquake 12 hours 4 minutes 30 seconds before its happening. However, the accuracy of the estimation of the magnitude of the earthquake is around 40 percent.

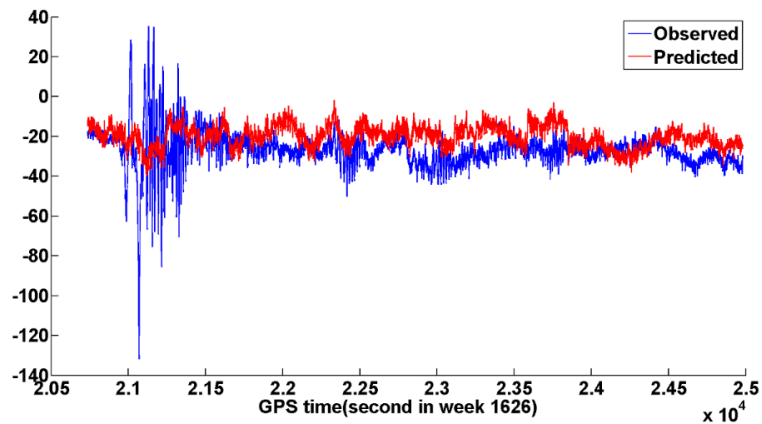

**Fig 7** Predicted earthquake in the station number 1 in the GEONET, *E* component, 10 percent of the data as training data

**Remark 7.** The predictability of the earthquake depends on the preseismic movements, hidden in the time series. The larger these movements, the more accurate the earthquake time and its measure of ground motion can be predicted. However, based on Section 2, it can be said that at least the 3 centimeters movements must be present in the training data.

**Conclusions**

A new machine learning algorithm is presented, which is specifically designed for the GNSS time series prediction. The algorithm is based on the characteristics of the GNSS time series. Its performance is checked against seventeen other conventional machine learning algorithms and one statistical method. In a large-scale study of 3000 GNSS time series from around the world, the proposed algorithm has a better accuracy than all the other algorithms, in addition to being faster. Applications of the proposed algorithm in the fields of geodesy and geodetic science are presented. The outlier and anomaly detection, and earthquake prediction problems are investigated. It is shown that the algorithm works well in these problems.

The algorithm in this paper is solely based on the GNSS time series. In the applications presented it would be of great use to combine the results of the proposed algorithm with other sources of information, such as satellite-based techniques. This paper does not deal with this problem. However, the promising results would be a motivation for the scientists working in the area of hazard assessment. For instance, the earthquake prediction problem can be pursued further by combining the geological data of the site, such as the temperature of the ground waters in the region (Orihara et al. 2014). However, as (Uyeda 2013) points out, the earthquake problem by this approach is often neglected. It would be ideal to use a combination of different data and models for this purpose, including the model in this paper.

**Data Availability**

The data in this paper are taken from the following sources
1. For the large-scale study and the outlier detection: http://geodesy.unr.edu
2. For the earthquake prediction: https://rdmc.nottingham.ac.uk/handle/internal/7006

**Author Biographies**

M. Kiani is an all-round geoscientist who works in all areas of geoscience, especially in geophysics and geodesy. His interests include both the theoretical and applied aspects of all areas of geosciences, among which are gravity field, physical, satellite, and mathematical geodesy, remote sensing, photogrammetry, hydrology, hydrography, GIS, geomorphology, and geology.